\documentclass[aps,preprint,superscriptaddress]{revtex4-1}

\usepackage[utf8]{inputenc}
\usepackage{mathtools}

\usepackage{graphicx}
\usepackage{dcolumn}
\usepackage{bm}
\usepackage{xcolor,colortbl}
\definecolor{Gray}{gray}{0.85}

\begin{document}
\title{Gender-specific behavior change following terror attacks}

\date{\today}
\author{Jonas S. Juul}
\affiliation{\mbox{Niels Bohr Institute, University of Copenhagen, Blegdamsvej $17$, Copenhagen 2100-DK, Denmark}}

\author{Laura Alessandretti}
\affiliation{\mbox{Centre for Social Data Science, University of Copenhagen, DK-1353 Kgs. København K, Denmark}}
\affiliation{Technical University of Denmark, DK-2800 Kgs. Lyngby, Denmark}

\author{Jesper Dammeyer}
\affiliation{Department of Psychology, University of Copenhagen, \O ster Farimagsgade 2A, 1353 K{\o}benhavn K, Denmark}

\author{Ingo Zettler}
\affiliation{Department of Psychology, University of Copenhagen, \O ster Farimagsgade 2A, 1353 K{\o}benhavn K, Denmark}

\author{Sune Lehmann}
\affiliation{\mbox{Centre for Social Data Science, University of Copenhagen, DK-1353 Kgs. København K, Denmark}}
\affiliation{Technical University of Denmark, DK-2800 Kgs. Lyngby, Denmark}

\author{Joachim Mathiesen}
\affiliation{\mbox{Niels Bohr Institute, University of Copenhagen, Blegdamsvej $17$, Copenhagen 2100-DK, Denmark}}

\begin{abstract}
Terrorists use violence in pursuit of political goals. While terror often has severe consequences for victims, it remains an open question how terror attacks affect the general population. We study the behavioral response of citizens of cities affected by $7$ different terror attacks. We compare real-time mobile communication patterns in the first $24$ hours following a terror attack to the corresponding patterns on days with no terror attack. On ordinary days, the group of female and male participants have different activity patterns. Following a terror attack, however, we observe a significant increase of the gender differences. Knowledge about citizens' behavior response patterns following terror attacks may have important implications for the public response during and after an attack.
\end{abstract}

\maketitle

\section{Introduction}
Terror attacks affect all parts of the world and are often carried out in attempts to communicate political messages or to dictate a political change~\cite{crenshaw1981causes,abrahms2006terrorism}. The European Parliament and Council of the European Union defines an offence as terror if it has one of three aims; the first of these is ``seriously intimidating a population''~\cite{eu-terror}.  While the direct consequences of an attack are easily quantified in terms of human casualties or material damage, the implications for the populations more broadly remains an open question. Of particular interest is how terror attacks impact the behavior of citizens. How do people react? Do people significantly change their behavior? If they do, is this change in behavior similar for all citizens, or are particular groups of individuals especially susceptible? Knowledge on such questions is sparse, yet valuable in making informed decisions about public response following terror attacks.

Previous studies have shown that exposure to terror increases the level of psychological stress as well as the frequency of disorders such as post traumatic stress, anxiety, and depression. For example, in the month following the 9/11 attacks, 12\% of the U.S. population experienced significant distress, about 30\% reported symptoms of anxiety and 27\% reported that they avoided situations that reminded them of 9/11 \cite{silver2002nationwide}. Exposure to terror through media or from knowing a victim also results in higher levels of avoidance behavior, a subjective feeling of insecurity, emotional distress, as well as changes of daily routines such as the choice of transportation \cite{korn2011affective, oppenheimer2011effects}. Whereas repeated exposure is known to be a physical health risk \cite{shenhar2015fear}, studies have shown that for isolated events, the frequency of stress symptoms quickly return to normal levels – for instance among citizens in New York City following the 9/11 terror attacks  \cite{schlenger2005psychological}. Individual differences, and to some extent gender differences, have been reported to be factors in the response to terror  \cite{gabriel2007psychopathological,galea2002psychological, kimhi2006women,salguero2011trajectory, pat2007risk}. For example, in a survey study women expressed symptoms of posttraumatic stress and depression more frequently than men \cite{solomon2005terror}. Women’s likelihood of developing posttraumatic stress symptoms were six times higher than those for men. 
It has also been reported that continuous threat of a terror attack promotes risk taking behaviors in men \cite{pat2007risk}. 

Objective real-time data have been used to analyze behavior patterns following e.g.\ natural disasters \cite{lu2012predictability,bengtsson2011improved}, emergencies \cite{bagrow2011collective}, and crowd disasters \cite{helbing2007dynamics,johansson2008crowd,bagrow2011collective}. Whereas objective data have been used in studies on the frequency and size of terror attacks~\cite{clauset2007frequency, clauset2013estimating} and structural properties of operational networks~\cite{clauset2008hierarchical,manrique2016women}, systematic studies on the behavioral impact of terror typically have relied on post-terror and self-report data. Consequently, quantitative studies of terror-related behavioral changes are much needed \cite{grimm2009risk}. In fact, the use of objective data to understand peoples response is limited to a few studies, for example in observations of correlations between the terror alert level and the number of people using public transportation \cite{montes2006effects},  in observations of an increase of fatal traffic accidents in the days following a terror attack \cite{stecklov2004terror}, and the behavioral response to a bombing with several injured and no fatalities~\cite{bagrow2011collective}. In the latter case, the authors found that females were more likely to make a call following the emergency, than expected on normal days. Here, we analyze behavior patterns in telecommunication following several recent terror attacks throughout Europe. 
We rigorously test for gender differences in the behavioral response to terror attacks and explicitly compare with ordinary days with no terror attacks.

\section{Results}
Our study uses data on telecommunication activity following 7 terror attacks in $6$ different cities: Paris, Nice, Berlin, London($\times 2$), Stockholm, and Barcelona. The attacks were carried out in the period November 14, 2015 - August 17, 2017, and although the attacks varied in size, all attacks resulted in several casualties (see Material and Methods).

\begin{figure}[t]
    \centering
    \includegraphics[width=5.0in]{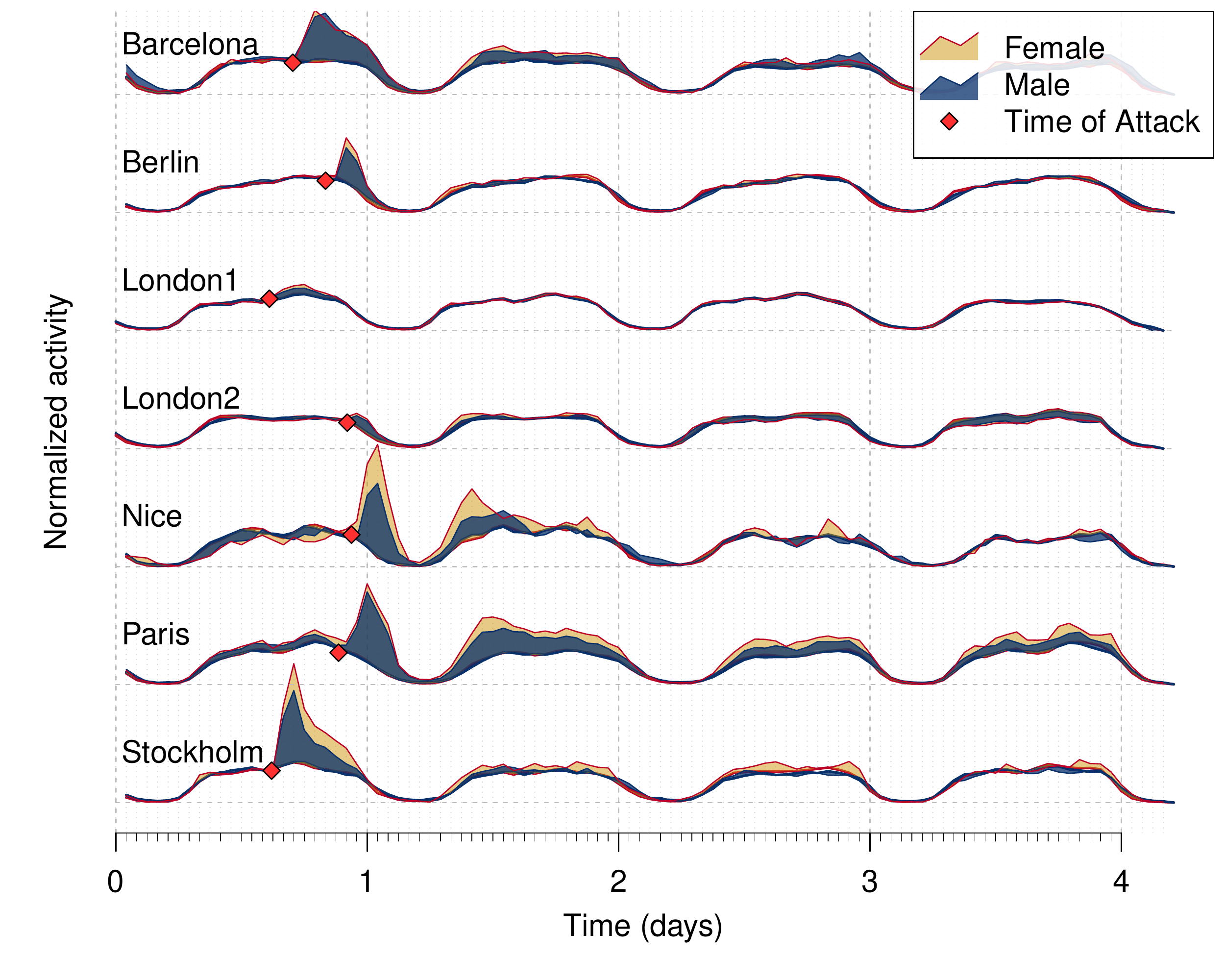}
    \caption{{\bf Normalized communication activity at the day of the terror attack and the three consecutive days.} The activity of the genders are normalized separately by their respective mean activities over $8$ background weeks. The colored area under the curves shows the increased activity relative to the background weeks.}
    \label{fig:separated_and_daily_activity}
\end{figure}

\begin{figure}[t]
    \centering
    \includegraphics[width=5.0in]{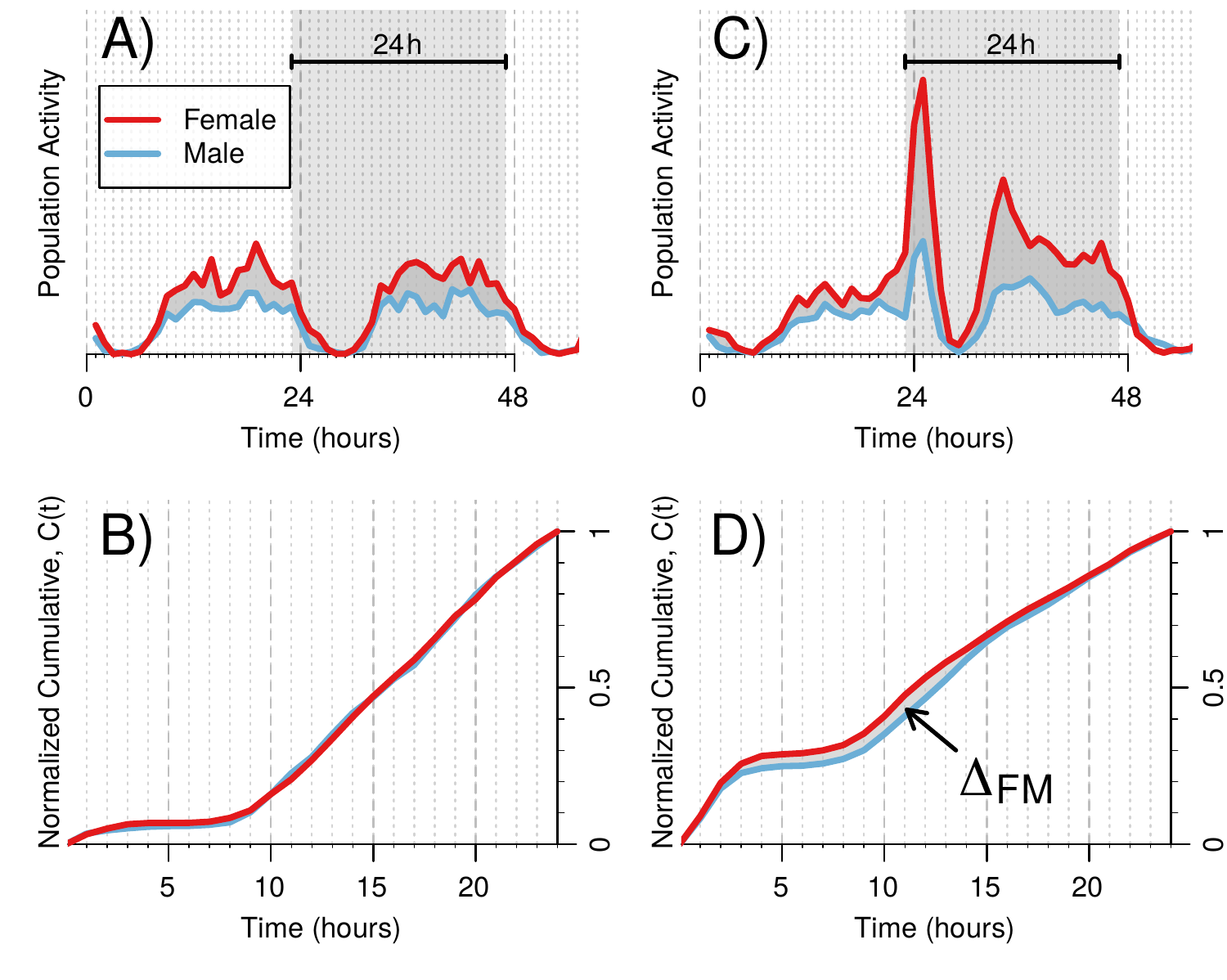}
    \caption{{\bf Comparison of gender activities following an attack and in a representative background week.} A) Example of activities on normal days and C) activity following a terror attack. For our analysis, we consider a $T_{max}=24hrs$ window. B) To quantify behavior differences between two groups of people, we use the area between normalized cumulative diurnal curves of telecommunication. On  normal days, our measure detects no difference in the relative gender activities, however, D) following an attack, we see a pronounced difference marked by the area $\Delta_{\alpha\beta}$. See Methods for technical details of how the measure can be computed from raw telecommunication data.}
    \label{fig:measure}
\end{figure}

Figure~\ref{fig:separated_and_daily_activity} shows that following a terror attack the activity deviates significantly from the normal diurnal rhythm~\cite{kondor2015visualizing,aledavood2015digital} (see Supplementary Section \ref{sec:diurnal}). In our analysis, we focus on the deviation in the $24$ hour window following terror attacks. We calculate the cumulative telecommunication activities for the female and male population separately, with notation $C_F(t)$ and $C_M(t)$, where the subscripts $F$ and $M$ refer to females and males, respectively. We normalize the cumulative activity such that for both populations, $C_{F/M}(0h)=0$ and $C_{F/M}(24h)=1$. We compute the area $\Delta_{FM}$ between the two curves, $C_F(t)$ and $C_M(t)$, and use this area to quantify the difference in behavior change (See Methods for details on our statistical analysis). Figure \ref{fig:measure} shows an example of the activity on an ordinary day (panel A) and on a day following a terror attack (panel C). The corresponding cumulative activities are shown in panels B and D. Note that in Fig. \ref{fig:measure}A, the telecommunication activity of females is higher. Averaging over all participants in our study, we find that females in general are $18\%$ more active than the male participants (see Supplementary Section \ref{sec:diurnal}). However, the normalized cumulative activity (Fig.\ \ref{fig:measure}B) shows that the only difference between the diurnal rhythm of the genders is the volume of the activity. Following a terror attack, however, the response of the female participants is significantly different to that of the males as illustrated in Fig.\ \ref{fig:measure}D and quantified by the area between the normalized cumulative activities.

Following each terror attack, we probe the difference between the gender specific communication patterns by computing the area $\Delta_{FM}$. We compare this area with a null distribution computed in the following way. First, we split the total population of both genders in two groups chosen at random: one group, $\tilde{F}$, which has a number of individuals equal to the number of females in the original population and a group $\tilde{M}$, consisting of the remaining individuals. Note that these two groups will contain a mix of both genders. Second, we compute an area $\Delta_{\tilde{F}\tilde{M}}$ between the normalized cumulative activities of the two new groups using the same 24 hour window following an attack. Repeating the process of randomly splitting our population $10^5$ times, we estimate the null distribution of $\Delta_{\tilde{F}\tilde{M}}$. 

\begin{figure}[t]
    \centering
    \includegraphics[width=3.5in]{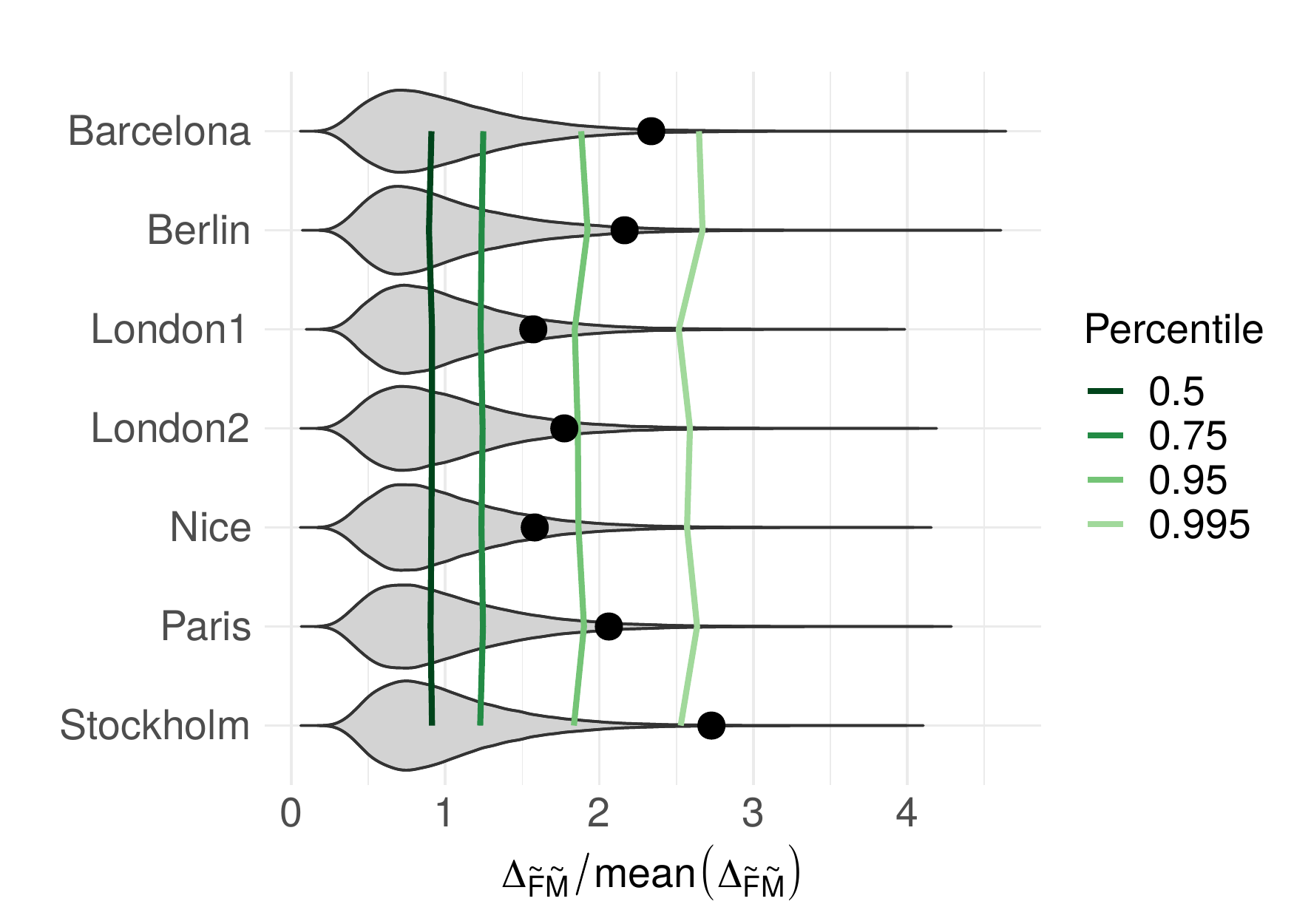}
    \caption{\textbf{Gender differences in telecommunication on days of terror attacks.} Empirically observed gender difference in telecommunication following terror attacks (black circles) plotted relative to computed null distributions (violin plot) for all cities. 
        The null-distributions were computed by randomly shuffling individuals between our two gender groups and measuring the difference in behavior of the new groups. Note that the true empirical values of the gender difference all lie beyond the $0.75$ percentile of the null-distribution.}
    \label{fig:gender_on_day_of_attack}
\end{figure}
We now test if the communication patterns of the female and male participants are significantly different on the day of the attack by computing the probability $p$ to observe an equally or more extreme value than the empirically computed value $\Delta_{MF}$, i.e.\ $p=$Prob($\Delta_{\tilde F\tilde M}\ge\Delta_{FM})$. Figure \ref{fig:gender_on_day_of_attack} shows the null distribution and corresponding values of $p$. Combining the measurements over all the cities (see Material and Methods) leaves that the gender-specific patterns are different with a combined $p$ value of less than $10^{-5}$.

Although the analysis above shows that the difference in behavior patterns is larger in the gender specific group than in randomly sampled groups, we cannot yet rule out that such a difference could be observed on ordinary days too. We therefore perform an additional test where we compare the behavior difference on the day of attack with the difference in the background weeks. Again, we quantify the behavior change by comparing $\Delta_{MF}$ to a null distribution of  $\Delta_{\tilde{M}\tilde{F}}$. Instead of shuffling gender labels (within the day of the attack), we now randomly choose a recorded activity of the individuals from the 8 background weeks. 
   Like above, we keep the sizes of the populations fixed, i.e. $|M|=|\tilde{M}|$ and $|F|=|\tilde{F}|$. 
We now compute a null distribution by replacing the 24 hour communication pattern of individuals with a randomly selected communication log (from a person of same gender) in one of the 8 background weeks. In this way, we keep the gender fixed, but the activity on the day of the terror is replaced by one from the background weeks. We finally quantify the difference in communication behavior by the area between the normalized cumulative diurnal telecommunication curves. We repeat this procedure $10^5$ times to get a null distribution of differences between males and females communication activity on ordinary days.

\begin{figure}[t]
    \centering
    \includegraphics[width=3.5in]{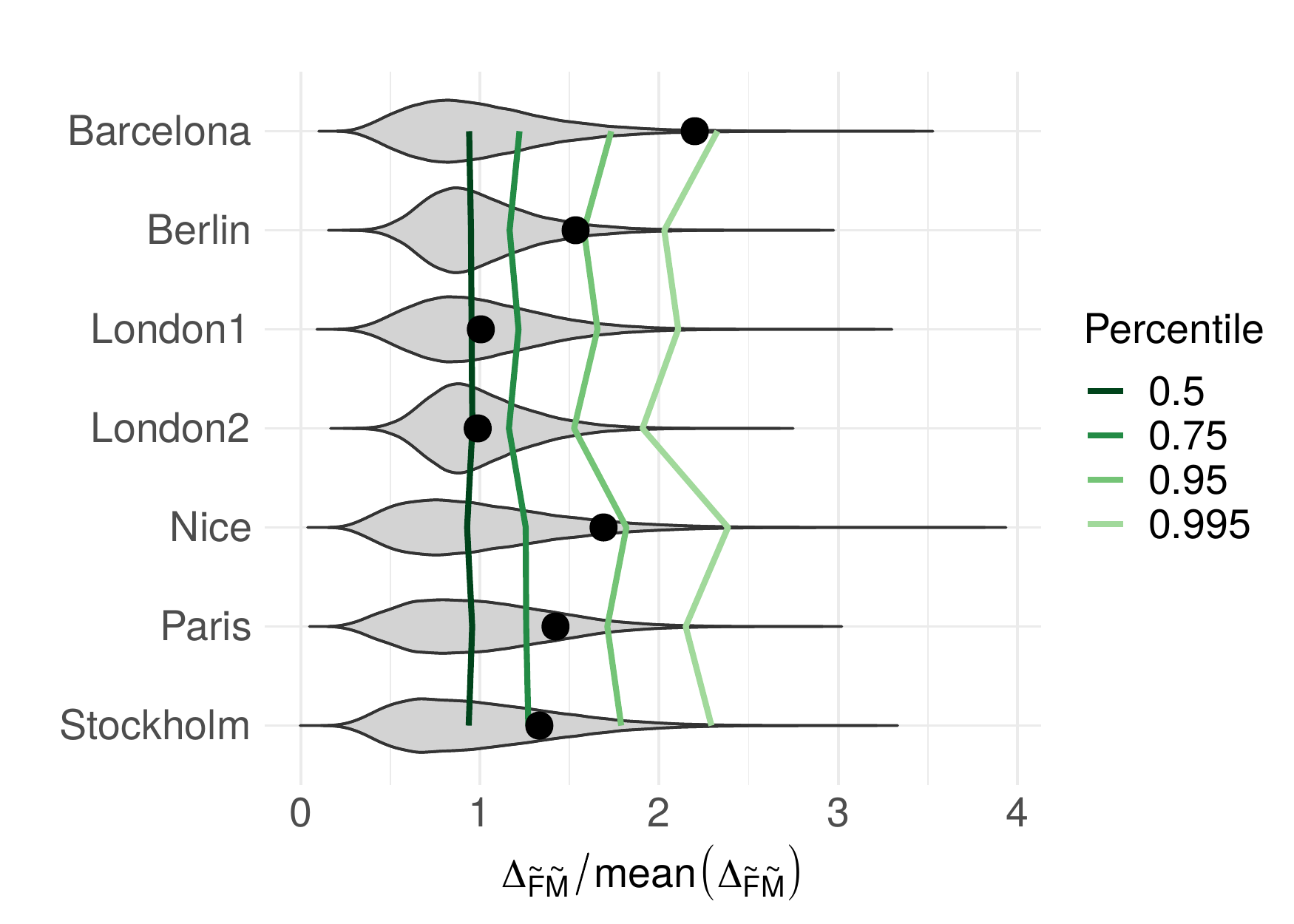}
    \caption{\textbf{Amplified gender differences in behavior following terror attacks.} 
    Empirically observed gender difference in telecommunication following terror attacks (black circles) plotted relative to computed null-distributions (violin plot) for all cities. 
    The null-distributions were created by computing the difference in telecommunication between groups of randomly selected males and randomly selected females on days with no terror. The random population of males (females) consisted of telecommunication logs drawn uniformly-at-random from the activity of all males (females) during the 8 background weeks. Note that the true empirical values of the gender difference all lie beyond the median of the null-distribution. The probability of getting a set of empirical values this or more extreme on days with no terror is approximately two in a thousand.}
    \label{fig:gender_comparison}
\end{figure}

We can now test if the differences between female and male communication patterns are more extreme on the day of the attack compared to ordinary days. To this end we compute the probability to observe an equally or more extreme value than the empirical $\Delta_{MF}$. Figure~\ref{fig:gender_comparison} shows our results. We find that, for all cities, this probability is smaller than $50\%$. Combining these probabilities yields a chance of getting a set of probabilities at least as extreme as these equal to ${p\approx0.002}$.

\section{Discussion}
Terror attacks take place around the world and it is important to understand how the general population reacts. Using real-time GPS and telecommunication data from $17,000$ people, we systematically studied behavior following $7$ different terror attacks in Western European countries. In particular, we studied the behavioral change of the groups of females and males in each city on the day of the attack as compared to ordinary days. We found that the telecommunication of both males and females spiked following all terror attacks. These spikes were significant deviations from normal telecommunication behavior. 

One might expect that a sizable disturbance of a population, such as a terror attack, might cause the distribution of calls of females and males to become more similar than they would be otherwise. Juxtaposing the behavior of the two genders following terror attacks, we found significant differences in the behavioral change of the two genders compared to a group where gender labels are assigned randomly. The differences between the groups of males and females after the attacks, were large even when correcting for differences on ordinary background days.

It is an open question what causes the observed gender differences in behavior following terror attacks. From a psychological perspective, differences in personality characteristics for males and females could be relevant. Previously reported differences between women and men in personality characteristics include scores in Emotionality and Honesty-Humility (for a recent meta-analysis, see for example \cite{moshagen2019meta}). In this light, it would be interesting to test whether the gender differences we observe can be attributed more directly to personality traits known to differ between genders. It is important to qualify the point that these results were obtained for the aggregated behavior of females and males. The uncovered gender differences on the aggregate level do not imply that every female and every male act significantly different from each other. The aggregated communication could be influenced by ``extreme'' individuals. If this is the case, such extremes seem to be present systematically in the different cities.  Furthermore, systematic studies of the variance in behavior patterns inside each group would be worthwhile to conduct.

Finally, future research should examine whether there are any cultural or regional differences in the behavioral response to terror attacks. In our analysis, we found significant gender differences in behavioral change following terror attacks supported by the analysis presented in Figure~\ref{fig:gender_comparison}. This figure also shows that although the difference in behavioral change is generally large, the empirical values for the two London attacks lie almost exactly at the null-distribution median. This begs the question whether London is different or if this is just random coincidence. If London is different, is this difference rooted in history, culture, geography, the nature of the attacks or some other variable? Knowledge on how terror attacks impact the general population is important in formulating a public response to such offenses, and we hope in this regard that our study will inspire further work on this topic.
\section{Methods}

\subsection{Experimental design}
We analyse behavioral change following $7$ terror attacks carried out in different European countries. Details of the different attacks are listed in Table~\ref{tab:attacks} including city name, time of the attack, the number of casualties and injured, and the number of people included in our data set. 
\begin{table}[]
    \centering
\begin{tabular}{lccc}
     \hline
     City & Time of attack & No. Fatalities/No. Injured & No. People in data \\
     \hline 
     \hline 
    \rowcolor{Gray}
    Paris (FR) & 00:58, November 14, 2015 & 137/368~\cite{europol2016te} & $2523$ \\
    Nice (FR) & 10:30 pm, July 14 2016  & 86/201~\cite{europol2017te} & $237$ \\
    \rowcolor{Gray}
    Berlin (DE) & 8:02 pm, December 19 2016 & 12/56~\cite{europol2017te} & $2295$\\
    London (UK) & 2:40 pm, March 22 2017 & 6/50~\cite{europol2018te} & $5415$\\
    \rowcolor{Gray}
    Stockholm (SE) & 2:53 pm, April 7, 2017 & 5/14~\cite{europol2018te} & 741\\
    London (UK) & 10:06 pm, June 3 2017 & 11/48~\cite{europol2018te} &$5131$ \\
    \rowcolor{Gray}
    Barcelona (ES) & 4:54 pm, August 17 2017 & 15/131~\cite{europol2018te} &$688$ \\
    \hline
\end{tabular}
    \caption{\textbf{Details about the terror attacks included in the study and the data concerning each attack.}}
    \label{tab:attacks}
\end{table}
We monitor the behavior difference using telecommunication data on the week of the attack and compare it with $8$ background weeks leading up to the terror attacks. For each background week, we consider the same $24$-hour interval as we do following a terror attack. The $8$ weeks used as background for each attack are listed in Table \ref{tab:background}.

\begin{table}[]
    \centering
    \begin{tabular}{l c c}
     \hline
     City & First background week Monday & Last background week Sunday \\
     \hline 
     \hline 
    \rowcolor{Gray}
    Paris (FR) & September $14$, 2015 & November $22$, 2015\\ 
    Nice (FR) & May $19$, 2016 & July $17$, 2016 \\
    \rowcolor{Gray}
    Berlin (DE) & October 24, 2016 & December $18$, 2016 \\
    London (UK) & January $25$, 2017 & March $21$, 2017\\
    \rowcolor{Gray}
    Stockholm (SE) & February $10$, 2017 & April $6$, 2017 \\
    London (UK) &  April $7$, 2017 & June $2$, 2017\\
    \rowcolor{Gray}
    Barcelona (ES) &  June $22$, 2017 & August $16$, 2017\\
    \hline
    \end{tabular}
    \caption{\textbf{Details about background weeks used in the analysis.}}
    \label{tab:background}
\end{table}
\subsection{Data description}
We used a dataset of phone-app usage and GPS records collected by a global mobile phone and electronic company between 2015 and 2017. We considered 7 terror attacks, and selected $\sim17,000$ users who lived in the same city where an attack occurred at the time it happened. Specifically, we selected users whose most visited location in the period under study (see Table \ref{tab:background}) is within a bounding box around the city where the attack happened (see Supplementary Table S1). We considered the usage of applications the company categorized as “Communication”. About 60\% of events in this category concern the usage of 5 Android apps: Phone, Messaging, WhatsApp Messenger, Facebook, Gmail and Facebook Messenger. Users are aged between 18 and 80 years old, with an average age of 36 years. About 42\% of the users are female. Written consent in electronic form has been obtained for all study participants.

\subsection{Statistical methods}

\subsubsection{Measure from telecommunication data}
In the following we formally define the measure \--- the area between normalized cumulative diurnal curves of telecommunication \--- we are use in our analyses. In a given time interval, $[0,t_{\rm max}]$, individual participants, $\gamma$, initiate $N_\gamma$ communication events at times $\{ t_{i,\gamma}\}_{i=1}^{N_\gamma}$. We define the activity function $A^{(\gamma)}(t)$ of an individual in terms of the point process 
\begin{equation}
    A^{(\gamma)}(t) = \sum_{i=1}^{N_\gamma}\delta(t-t_{i,\gamma}),
\end{equation}
where $\delta(x)$ is the Dirac delta function. The activity function of a population $X$ is the sum of individual activity functions, 
\begin{equation}
A_X(t) = \sum_{\gamma\in X} A^{(\gamma)}(t).
\label{eq:population_activity}
\end{equation}
For each population activity function $A_X(t)$, we define the corresponding normalized cumulative activity function
\begin{equation}
    C_X(t) = \frac{\int_0^tA_X(t')dt'}{\int_0^{t_{\rm max}}A_X(t')dt'}.
    \label{eq:C}
\end{equation}
where the denominator is the total number of initiated communication events in our population, $N_X=\sum_{\gamma\in X} N_\gamma$, and thus $C_X(t)$ is equal to the fraction of communication events that were initiated before the time $t$. To assess the differences in communication patterns for two populations $X$ and $Y$ in a time interval $[0,t_{\rm max}]$, we compare how communication events are distributed over the time interval for the two populations. Specifically, we calculate the area between the cumulative activity functions for the two populations,
\begin{equation}
    \Delta_{XY} = \int_0^{t_{\rm max}}\left|C_X(t)-C_Y(t)\right|dt.
    \label{eq:measure}
\end{equation}
In this study, the length of the time interval, $t_{\rm max}$, is fixed to be $24$ hours. On a normal day, females, in our population, are on average 18\% more active than males. Fig.\ \ref{fig:measure}A illustrates this; similar curves are shown for the all cities included in this study in Supplementary Figure 1. The measure, defined in Eq.~\eqref{eq:measure}, has a number of attractive features. It is not sensitive to the imbalance in gender activities and allows us to quantify changes to the diurnal rhythm. Moreover, our measure does not require any artificial binning of data and is not particularly sensitive to the chosen time interval. 

\textit{Combining the results to probe for gender differences}. In our analysis, we obtain probabilities for the random occurrence of gender differences of the same size as observed following the terror attacks. If the null hypothesis, that there are no enhanced gender differences, were true, these probabilities would be uniformly sampled on the interval $[0,1]$. We test this as follows. If $X_1, X_2,\ldots, X_n$ are stochastic variables drawn from a uniform distribution, the random variables
\begin{equation}
    Y_i = -2\ln(X_i),
    \label{eq:Yi}
\end{equation}
are independent and identically distributed according to the chi-square distribution with $2$ degrees of freedom. The sum of these variables
\begin{equation}
    T = \sum_{i=1}^n Y_i,
    \label{eq:sum_Yi}
\end{equation}
is distributed according to the chi-square distribution with $2n$ degrees of freedom. To test whether our obtained probabilities support the hypothesis that the gender differences are not larger than should be expected from ordinary days, we calculate this $T$ value using Equations \eqref{eq:Yi} and \eqref{eq:sum_Yi}. We then calculate the area under the chi-square distribution with $2n$ degrees of freedom, at values larger than $T$,
\begin{equation}
    p_{\rm combined} = \int_{T}^\infty \chi^2_{2n}(x) dx.
    \label{eq:pcombined}
\end{equation}
This integral is equal to the chance of getting the set of probabilities if $X_1,X_2,\ldots,X_n$ were drawn from a uniform distribution.
\subsubsection{$p$-values used in calculating probability of observing behavior differences randomly}
The distributions depicted in Figures~\ref{fig:gender_comparison} and \ref{fig:gender_on_day_of_attack} give us two sets of $p$-values. In each case, we used the method described above to estimate the probability that the empirically observed measure values would be obtained if the values were results of statistical fluctuations. The $p$-values are shown in Tab.~\ref{tab:p_values_violins}. Combining these $p$-values give the combined probabilities listed in the main text.
\begin{table}
    \centering
    \begin{tabular}{l c c c c c c c}
    \hline 
        & \textbf{Barcelona} & \textbf{Berlin} & \textbf{London1} & \textbf{London2} & \textbf{Nice} & \textbf{Paris} & \textbf{Stockholm}\\
        \hline \hline
            \rowcolor{Gray}
\textbf{Fig~\ref{fig:gender_on_day_of_attack}} & $0.01312$ & $0.02543$ & $0.10505$ & $0.06394$ & $0.10809$ & $0.03104$ & $0.00233$\\ 
\textbf{Fig~\ref{fig:gender_comparison}} & $0.00862$ & $0.06071$ & $0.44336$ & $0.45604$ & $0.07581$ & $0.15288$ & $0.21411$\\
\hline
    \end{tabular}
    \caption{\textbf{Fraction of null distributions more or equally extreme as empirical values in Figs.~\ref{fig:gender_on_day_of_attack} and \ref{fig:gender_comparison}}. The fraction of null distributions depicted in Figs.~\ref{fig:gender_on_day_of_attack} and \ref{fig:gender_comparison} that lie beyond the corresponding empirically observed values. We subtract each of these values from $1$ and plug them into Eq.~\eqref{eq:Yi} and then \eqref{eq:sum_Yi} to calculate a combined probability using Eq.~\eqref{eq:pcombined}.}
    \label{tab:p_values_violins}
\end{table}


\bibliographystyle{unsrt}
\bibliography{terror_bibliography.bib}
\textbf{Acknowledgments:} 
\textbf{Funding:} J.S.J. J.D. I.Z. S.L. and J.M. received funding through the University of Copenhagen UCPH 2016 Excellence Programme for Interdisciplinary Research.
\pagebreak

\newcommand{\beginsupplement}{%
        \setcounter{table}{0}
        \renewcommand{\thetable}{S\arabic{table}}%
        \setcounter{figure}{0}
        \renewcommand{\thefigure}{S\arabic{figure}}%
        \setcounter{section}{0}
        \renewcommand{\thesection}{S\Roman{section}}
     }
     
\beginsupplement

\section*{\Large Supplementary Materials}

\section{Geography of the cities under study.}

We selected individuals whose most visited locations during the period under study (see Table~\ref{tab:background}) is located within the city where the attack happened. The bounding boxes characterizing each city are described in Table~\ref{tab:boxes}.

\begin{table}
    \centering
    \begin{tabular}{p{2.5cm} p{2.5cm} p{2.5cm} p{2.5cm} p{2.5cm}}
    \hline 
        city & min lat & max lat & min lon & max lon \\
        \hline 
        \hline 
            \rowcolor{Gray}
        Berlin &  52.369276 & 52.650018 & 13.091432 & 13.754525 \\
        Nice & 43.646275 & 43.758400 &  7.178630 &  7.338724 \\
        \rowcolor{Gray}
        Barcelona & 41.310933 & 41.465339 & 2.058793 & 2.244023 \\
        London & 51.325628 & 51.672014 &  -0.472381 & 0.268712 \\
        \rowcolor{Gray}        
        Stockholm & 59.298186 & 59.371545 & 17.945337  & 18.154841\\
        Copenhagen & 55.5531 & 55.8175 & 12.2607 & 12.7043 \\
        \rowcolor{Gray}  
        Paris & 48.7106 & 48.9991 & 2.0641 & 2.6463 \\
        \hline
    \end{tabular}
    \caption{Bounding boxes used to select individuals living in the cities under study. The table reports the minimum and maximum values of the latitude and longitude. }
    \label{tab:boxes}
\end{table}

\section{Diurnal communication on ordinary days}
\label{sec:diurnal}
\begin{figure}[ht]
    \centering
    \includegraphics[width=5.0in]{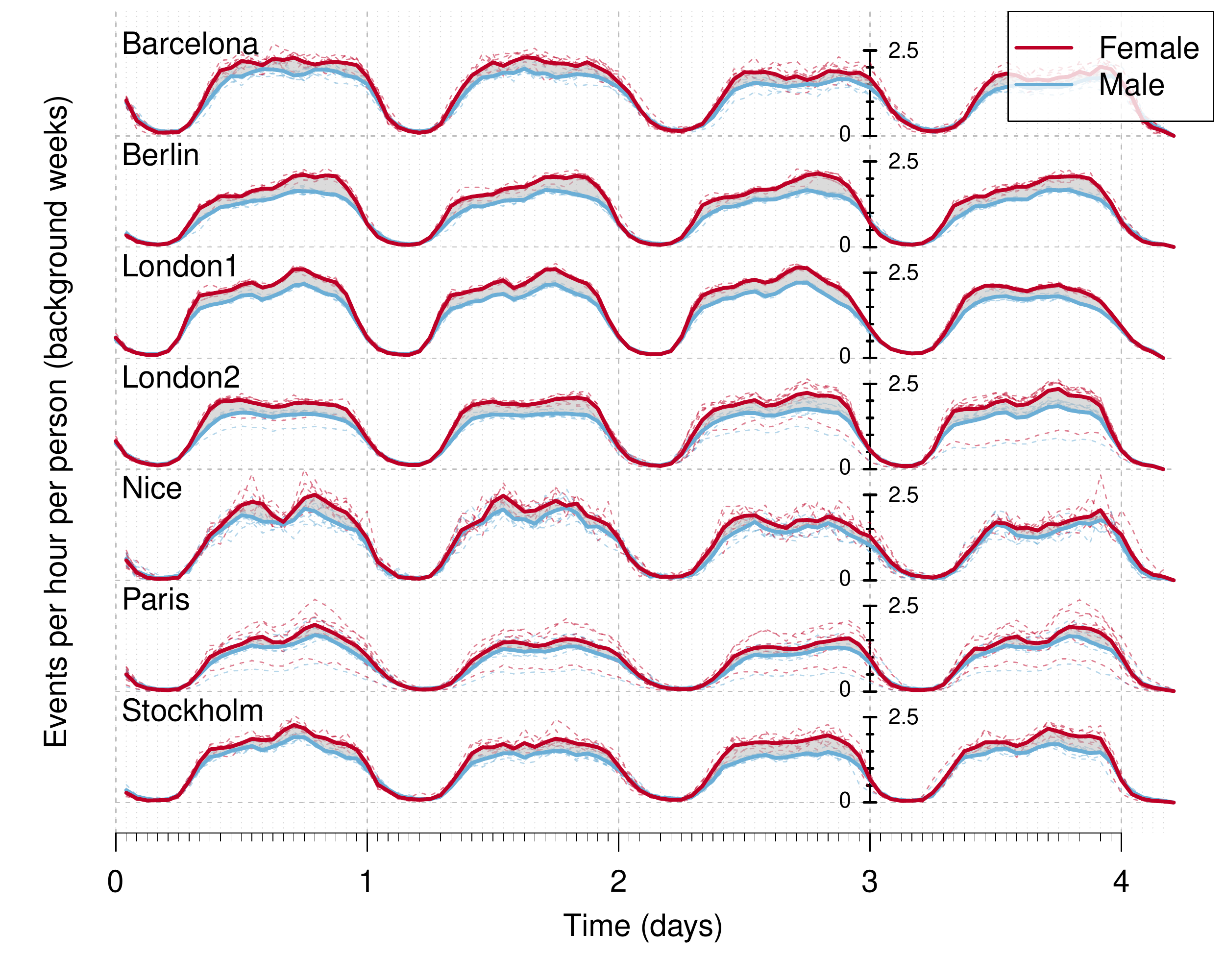}
    \caption{Illustration of diurnal communication patterns during ordinary days. Data for all background week are plotted in dashed lines. The mean activity for each gender is plotted with full lines. On average, females communicate $18\%$ more than men.}
    \label{fig:gender_activity}
\end{figure}

The populations of males and females show distinct diurnal communication patterns in all of the cities we include in our analysis. Fig~\ref{fig:gender_activity} shows these curves for all background weeks used in our study. The curves are almost identical for all background weeks, indicating that averaging over the weeks yields a good estimate of the diurnal activity. Females communicate on average $18\%$ more than males but normalizing their activity yield indistinguishable curves. These normalized curves are shown in Fig.~\ref{fig:separated_and_daily_activity} of the main text. 
\section{Significance of peaks following terror attacks}
Figure~\ref{fig:separated_and_daily_activity} shows that a terror attack is followed by a spike in telecommunication activity for both genders for all cities in our study. 
In order to assess the significance of the spikes in the communication, we compare with a null model in the following way. We first compute a null-distribution of the area between normalized cumulative diurnal telecommunication curves by bootstrapping. The null-distribution quantifies natural variation to the diurnal pattern. More specifically, in a population of for example $n$ males, we create a set of individual activities on ordinary days $\{A^{(p)}\}_{p\in M}$ where $M$ is the subset of males in our population and each element is a $24$ hour sequence of communication events, see Eq.\ \eqref{eq:population_activity}. This set has $8n$ elements, one for each person and for each of the 8 background weeks. We draw $n$ random elements from $\{A^{(p)}\}_{p\in M}$ (allowing for repeated draws of the same element) corresponding to $n$ activity functions, from which we get a single background cumulative activity function using Eqs.~\eqref{eq:population_activity} and \eqref{eq:C}. We then choose one of the $8$ ordinary days, $i$, and calculate the area between the cumulative activity functions for the male population at day $i$, and the cumulative activity function for the $n$ randomly picked individuals. By repeating this procedure $10^5$ times for each background week $i$, we obtain $8$ null distributions of areas.

Then, we test the alternatives (a) and (b), by computing the probability to observe the empirical communication activity measured on the day of the attack during any of the background weeks. We obtain these empirical values by calculating the area between the cumulative activity function of the male population on the day of the attack, and the cumulative activity function of the male population on each of the ordinary days. The percentage of the measure-value null distribution of week $i$ that is larger than or equal to the empirical value for week $i$ represents the probability that the empirical value is a result of 
random noise. These $8$ probabilities can be combined into a single $p$-value, expressing the likelihood of getting the $8$ probabilities given that the telecommunication was unaffected compared to ordinary days (see Methods). For all cities and both genders, we find that the probability associated to alternative (a) is less than $1.88\cdot10^{-5}$, revealing that the telecommunication observed in the $24$ hours following the terror attacks is unlikely to be observed on ordinary days (see Table~\ref{tab:combined_pvalues_spikes}).

The least significant peak is the one exhibited by females after the second London attack. For this population, we obtained the p-values $0.99988$, $0.8216$, $0.89666$, $0.9994$, and $0.99502$. Combining these in the way described in Methods yields the single p-value $1.88\cdot10^{-5}$. Other than this population, only $4$ populations did not have every p-value smaller than $10^{-5}$. The combined p-values for these are all below $10^{-5}$ (best estimate is listed in Table~\ref{tab:combined_pvalues_spikes}).
\begin{table}
    \centering
    \begin{tabular}{l c c}
    \hline 
        City & Gender & Combined \\
        \hline 
        \hline 
            \rowcolor{Gray}
        Nice & Male & $2.91\cdot10^{-19}$ \\
        London2 & Female & $1.88\cdot10^{-5}$ \\
        \rowcolor{Gray}
        London2 & Male & $7.59\cdot10^{-6}$ \\
        London1 &  Female & $7.08\cdot10^{-20}$\\
        \rowcolor{Gray}        
        London1 & Male & $1.07\cdot10^{-36}$ \\
        \hline
    \end{tabular}
    \caption{Combined $p$-values for peaks following terror attacks (peaks illustrated in Fig.~\ref{fig:separated_and_daily_activity}). All combinations of attack and gender that are not listed had $p$-values indistinguishable from $0$.}
    \label{tab:combined_pvalues_spikes}
\end{table}
\end{document}